\documentstyle [12pt] {article}

\begin{document}
\title {\large \bf Q-BOSON REPRESENTATION OF THE QUANTUM MATRIX
ALGEBRA\ \  $M_q(3)$ }
\author{Vahid Karimipour}
\date { }
\maketitle
\begin{center}
{\it Institute for Studies in Theoretical Physics and Mathematics \\
P.O.Box 19395-1795, Tehran, Iran\\ Department of Physics, Sharif
University of Technology \\
P.O.Box 11365-9161, Tehran, Iran}\\
\end{center}

\vspace {10 mm}
\begin {abstract}
{Although q-oscillators have been used extensively for realization of
 quantum
universal enveloping algebras,such realization do not exist for quantum
matrix
algebras ( deformation of the algebra of functions on the group ). In this
paper we first construct an infinite dimensional representation of the
 quantum
matrix algebra $ M_q ( 3 ) $(the coordinate ring of $ GL_q (3)) $ and
then use this
representation to realize $ GL_q ( 3 ) $ by q-bosons.}
\end{abstract}
\noindent
\newpage
{\large \bf I. Introduction}\\

Since the advent of q-oscillators or q-boson algebras [ 1-3 ]  a lot of
attention
has been paid to realization of quantum universal enveloping
algebras ( QUEA )
[4-8] in terms of q-oscillators.
However the corresponding task for the dual objects , i.e: quantum matrix
algebras, has not been
studied so far , except for the case of $ GL_q (2 ) $ [ 9 ]. In this
paper we
extend the results of [9]
and give a 3-parameter family of q-boson realization for the quantum matrix
algebra $ M_q ( 3 ) $.

The quantum algebra $ GL_q(3)$ is genereated by the elements of a matrix
\begin{equation} T = \left(\begin{array}{lll} a & b & c   \\ d & e & f
\\ g & h & k  \end{array}
\right)
\end{equation}
subject to the relations:
\begin{equation}
R  T_1 T_2 = T_2 T_1 R
\end{equation}
where R  is the solution of the Yang Baxter equation corresponding to $ SL_q
(3) $.[5].
\begin{equation}  R  =\sum _{i\ne j}^n e_{ii} \otimes e_{jj} +\sum _
{i=1}^n  q
e_{ii}\otimes e_{ii} +(q -q^{-1}) \sum _{i<j}^n e_{ji}\otimes e_{ij}
\end{equation}
The relations
obtained from (2) can be expressed neatly in the following form :
For any $ 2\times 2 $ submatrix ( i.e : like the one formed by the elements
$ b , c, e, $ and $ f $ )
 the folllowing relations hold:

$$ bc = q cb \hskip 1cm ef = q fe  $$
\begin{equation}
be = q eb \hskip 1cm cf = q fc
\end{equation}
$$ ec = ce \hskip 1cm bf - fb = (q - q ^{-1}) ce $$\\
{\bf REMARK}: These relations are only a small part of the relations
obtained
from
(2).All the other relations can be simply read by looking at other
submatrices.( i.e: $ df = q fd\ \  , dg = q gd\ \  , dk - kd =
( q - q^{-1} ) fg\ \  , $  etc
). Hereafter when we refer to ( 4 ) we mean all the relations of
which ( 4 ) is
a sample.
Thus this algebra has  many $ GL_q(2) $ subalgebras ( i.e: the
elements
$ a,\  c,\  d,\  and \ \ f $ generate a $ GL_q(2) $ subalgebra ).Obviou
sly these are not
Hopf subalgebras.\\ \\
{\bf REMARK}:
One can prove the following  more general
type of formula
\begin{equation}
bf^n - f^n b = q^{-1} ( q^{2n} -1 ) f^{n-1} ce
\end{equation}
by induction from (4).

$ GL_q(3) $ has also a quantum determinant D [ 8 ] which is central:
\begin{equation}
D = a \Delta_a - q\  b \ \Delta_b + q^2\   c\  \Delta _c
\end{equation}
where $ \Delta_a  \ \ \Delta_b  $ and $ \Delta_c $ are the quantum cofac
tors of the elements $ a \ \ \ b  $ and $ c $ respectively :
\begin{equation}
\Delta_a = ek - q fh \hskip 2cm \Delta_b = dk - q fg \hskip 2cm
\Delta_c = dh - q eg
\end{equation}
One can also see from (4) that the elements $ c \  \ e $  and $ g $
commute with each other,a fact which
will play an important role in building up the representation.
It is clear that the eigenvalues of the operators $ c , e  $ and $ g $
will label the states of any representations. What remains to be
done is the choice of
lowering and raising operators. At first sight one may try to choose
the operators
$ f , h $ and $ k $ as raising and $ a , b $ and $ d $  as lowering
operators , and construct
a Verma module out of the states\ \  $  \vert l,m,n > \equiv f^l\
h^m\  k^n \vert 0 >
$ \ \ where \ \  $ \vert 0 > $ \ \ is the vaccum which  is an eigenstate
 of $ c ,
e $\   and \ $ g $\   and satisfies:\ \ $ a\ \vert 0 > \ =\  b\ \vert
 0 > \ = \ d\ \vert 0 > \ = 0
$
\ \ But this choice has the disadvantage that to compute the action of
a lowering operator
like $ a  $ on $ \vert l,m,n> $  one must  use the commutation
relations of the type (5) many times
which makes the computation cumbersome and the results not
 illuminating. However
a much better approch is possible , which we now explain.\\ \\
{\large \bf II. Infinite Dimensional Representation }\\

We first construct an
infinite dimensioal representation of a subalgebra of $ M_q (3) $. This
subalgebra  which we will denote by $ A $ is
generated\ \   D \ \ and all the elements of the matrix $ T $  ( except
 $ a $  and  $ k $ ) plus two quantum
cofactors $ \Delta_a $  and  $ \Delta_ k  $ .Hereafter we denote them by $
\Delta $ and $ \Delta' $ respectively .
\begin{equation} \Delta = ek - q fh \hskip 2cm \Delta' = ae - q bd
 \end{equation}
As we will see the elements $ \Delta $ and $ \Delta' $ rather than $ k $
 and $ a $  will be the natural choice of the third pair
of raising and lowering operators. The first two pairs are ( f , b ) and
 ( h ,d )
Clearly the element $ \Delta $  being the q-determinant of the submatrix
$  \left( \begin{array}{ll} e & f  \\ h & k \end{array} \right)     $
commutes with
 the elements $ e, f, h $  and  $ k $ . A similar statement holds true for
  $ \Delta' $ . ( i.e:$ \Delta' $ commutes
with $ a , b, d $ and $ e $  ) .
Using (4,8) it is straightforward to verify the following commutation
relations: $$ b \Delta  = q \Delta b \hskip 3 cm c \Delta  = q \Delta c $$
\begin{equation} d \Delta  = q \Delta d \hskip 3 cm
g\Delta  = q \Delta g  \end{equation}
and
$$ c \Delta'  = q^{-1} \Delta' c \hskip 3 cm g \Delta'  = q^{-1}
\Delta' g $$
\begin{equation} f \Delta '  = q^{-1} \Delta ' f \hskip 3 cm h \Delta
 '  = q^{-1}  \Delta ' h
\end{equation} We need two other relations which we present in :\\ \\
\setcounter{equation}{11}
\setcounter{equation}{12}
{\bf Lemma}:\\

i) $ a \Delta = q^2 \Delta a + ( 1-q^2 ) D \hskip 8cm \ \ \ \ \ \ \ \
\ \ \ \ \
\ \ \ \ (11)$
ii) $ \Delta'  \Delta = q^2 \Delta \Delta'  + ( 1-q^2 ) De \hskip
7cm\ \ \ \ \
\ \ \ \ (12) $\\
Where $ D $ is the quantum determinant of the matrix T ( see (6) ) .\\ \\
{\bf Proof} :

i) Passing $ a $ through $ \Delta $  and using the commutation relations
(4)
we find
$$ a \Delta = \Delta a + ( q - q^{-1} ) ( ecg + bdk - q fbg - q cdh ) $$
from (4) we have:
$$ bdk - q fbg = bdk - q ( bf - ( q-q^{-1})ce ) g = b\Delta_b +
( q^2 - 1 ) ceg
$$
therefore the sum of the terms in the bracket is equal to $ b \Delta_b -q
c \Delta_c $  and hence the above relation is transformed to the
following form: $$ a\Delta = \Delta a + (q-q^{-1})( b \Delta _b
- q c \Delta _c ) $$
By using the expression of the quantum determinant (6) we finally arrive at
( 11 ).

The proof of ii) is straightforward. One only needs the result of part i)
 and
eqs. (9-10).\\ \\
\setcounter{equation}{13}
\setcounter{equation}{14}
{\bf Corralary}:\\

i) $ a \Delta^n = q^{2n } \Delta^n a + ( 1-q^{2n}) D \Delta^{n-1}
\hskip 5cm \ \
\ \ \ \ \ \ \  $

ii) $ \Delta ' \Delta^n = q^{2n} \Delta^n \Delta' + ( 1-q^{2n}) D e
\Delta^{n-1} \hskip 4cm \ \ \ \ \ \ \ \ \ \ \ \  $\\
These formulas are proved by induction from formulas (11) and (12) .
One must use the commutativity of $ e $ and $ \Delta $ and the fact
that\ \ \ $ D $  is central.
We now construct an infinite dimensional representation of A .
 Let us denote by $ \vert 0 > $  a common eigenvector of $ c , e, $
and $ g $
which is annihilated by the lowering operators:
\begin{equation} b\vert 0 > =  d\vert 0 > =   \Delta' \vert 0 > = 0
\end{equation}
\begin{equation} c\vert 0 > = \lambda \vert 0 > \hskip 1cm
e\vert 0 > = \mu \vert
0 > \hskip 1cm    g \vert 0 > = \nu \vert 0 >  \end{equation}
Then we construct the q-analogue of Verma module as follows:
\begin{equation} W = \{ \vert l,m,n> \equiv f^l h^m \Delta ^n \vert 0 >
\hskip 1 cm l ,
m , n \geq 0 \}\end{equation}
The vectors of this Verma module are eigenvectors of $ c , e $ and $ g $.
$$ c\vert l,m,n> = q^{l+n}\lambda  \vert l,m,n> $$
\begin{equation} e\vert l,m,n> = q^{l+m}\mu \vert l,m,n> \end{equation}
$$ g\vert l,m,n> = q^{m+n}\nu \vert l,m,n> $$
Since $ f $ , $ h $ and $ \Delta $ commute with each other we obtain :
$$ f\vert l,m,n> =  \vert l+1,m,n>$$
\begin{equation} h\vert l,m,n> = \vert l,m+1,n>  \end{equation}
$$ \Delta\vert l,m,n> = \vert l,m,n+1> $$
\setcounter{equation}{20}
The action of the lowering operators are determined by using (9-10).
The result
is presented in :\\ \\
{\bf Theorem }:\\

i) $ b\vert l,m,n> = q^{m+n-1} ( q^{2l} - 1 ) \lambda \mu
 \vert l-1 , m , n > $

ii) $ d\vert l,m,n> = q^{l+n-1} ( q^{2m} - 1 ) \mu \nu \vert l , m-1 , n >
\hskip 5cm \ \ \ \ \ (20) $

iii) $ \Delta'\vert l,m,n> = q^{l+m } (1- q^{2n} )  \mu \eta
\vert l , m , n-1 > $

where $ \eta $ is the value of the quantum determinant in the
representation
$ D \vert l,m,n > = \eta \vert l,m,n > $.\\ \\
{\bf Proof}: We only prove  iii) . The other two parts are similar.

{}From (9,10) we have
$$ \Delta' \vert l,m,n> = \Delta' f^lh^m \Delta^n\vert 0 > =
q^{l+m} f^l h^m
\Delta'\Delta^n \vert 0 > $$
$$=q^{l+m} f^l h^m  ( q^{2n}\Delta^n\Delta'+ (1-q^{2n})De\Delta^{n-1})
\vert
0 > $$ $$ = q^{l+m}( 1-q^{2n})\mu \eta \vert l,m,n-1> $$

where we have used the commutativity of $ e $ and $ \Delta $ and the
centrality of $ D $.Eqs. ( 18-20 ) show that W is an infinite
dimensional A module.
Once one has a representation  of the subalgebra $ A $ , it is then
an easy task to determine
the representation of $ M_q ( 3 ) $ itself. One must only determine the
action of the generators $ a $ and $ k $  on $ \vert l,m,n> $.\\ \\
\setcounter{equation}{21}
\setcounter{equation}{22}
{\bf Theorem }:\\

i) $ k\vert l,m,n> = q^{-(l+m)} \mu^{-1} ( \vert l,m,n+1> +
q^{-1} \vert l+1 ,
m+1, n> )  \hskip 3cm (21)$

ii) $ a\vert l,m,n> = ( 1-q^{2n})\eta\vert l,m,n-1> + \mu \nu
\lambda ( q^{2m}-1)
( q^{2l}-1)q^{2n-2}\vert l-1,m-1,n> \  (22)$\\ \\
{\bf Proof} :  From (19) we have :
$$ \Delta \vert l,m,n> = \vert l,m,n+1 > $$
Using the fact  that $ \Delta $ has an equivalent description , namely $ \
Delta = ke - q^{-1} fh $ we find :
$$ ( ke-q^{-1}fh) \vert l,m,n> = \vert l,m,n+1> $$                        or
$$ q^{l+m}\mu k\vert l,m,n> - q^{-1} \vert l+1,m+1,n> = \vert l,m,n+1> $$
which proves i) .

For ii) we use a similar method :
$$ \Delta' \vert l,m,n> = ( ae - qbd) \vert l,m,n> $$
from which we obtain using ( 20 ):
$$ q^{l+m} ( 1-q^{2n})\mu \eta\vert l,m,n-1> = q^{l+m} \mu a\vert l,m,n>$$
$$ - q ( q^{l+n-1} ( q^{2m}-1) \mu \nu q^{m+n-2} (q^{2l}-1)\
lambda\mu \vert l-1,m-1,n> ) $$
Rerarranging the terms gives (22) .\\ \\
{\bf Remark }:  $ \eta $ is not an independent parameter.One can
determine its
value
by acting on any state with $ D $. The result is
$ \eta = - q^{-3} \lambda \mu \nu $ .\\ \\
{\large \bf III. Q-Boson Realization of $ M_q ( 3 ) $}\\

Having constructed the infinite dimensional representation we are now re
ady to realize the generators of  the quantum matrix
group by q- bosons. We proceed along the lines proposed in [9] .
 Let us denote
by $ B_q $ the q- Boson algebra generated by elements
$ a $ and $ a^{\dagger } $ with the relations:
$$ a a^{\dagger} - q^{\pm 1}a^{\dagger} a = q^{\mp N }$$
\begin{equation} q^{\pm N} a = q^{\mp 1} a q^{\pm N }\end{equation}
$$ q^{\pm N} a^{\dagger } = q^{\pm 1} a^{\dagger }  q^{\pm N } $$

We consider the algebra $ B_q^{\otimes 3 } $ generated by 3 commuting
 q-bosons
and its natural representation on the q-Fock space $ F_q $
$$ \vert \vert l,m,n>> \equiv (a_1^{\dagger})^l (a_2^{\dagger})^m
(a_3^{\dagger})^n \vert 0 >> $$\\
$$ a_1\  \vert\vert l, m, n >> = [l]_q\ \  \vert\vert l-1,m,n>>  , \ $$
\begin{equation} a_2\  \vert\vert l, m, n >> = [m]_q\ \  \vert\vert
 l,m-1,n>>  \end{equation}
$$ a_3\  \vert\vert l, m, n >> = [n]_q\ \  \vert\vert l,m,n-1>>  , \  $$

$$ q^{N_i } \vert \vert 0 >> = \vert \vert 0 >> \hskip 2cm i = 1, 2, 3, $$

If $ \Psi $ is the natural isomorphism from W to $ F_q $ satisfying:
$$ \Psi : \vert l, m, n > \longrightarrow \vert \vert l, m, n >> $$
then the induced representation $ \Psi $ is defined by :
\begin{equation} \Psi^* (g ) = \Psi \circ g \circ \Psi^{-1}  \hskip 2
cm \forall g \in End \
\ W\end{equation}
We will then find the following 3-parameter family of q-boson real
ization of
$ A $ .

$$ f \equiv a_1^{\dagger } \hskip 2cm h \equiv   a_2^{\dagger } \hskip 2cm
  \Delta \equiv  a_3^{\dagger }  $$
$$ c \equiv \lambda q^{N_1 + N_3 } \hskip 2cm    e \equiv \mu q^{N_1 +
 N_2 } \hskip 2cm  g \equiv \nu q^{N_2 + N_3 } $$
\begin{equation} b \equiv ( q-q^{-1}) \lambda \mu q^N a_1 \end{equation}
$$ d \equiv ( q-q^{-1})  \mu\nu q^N a_2 $$
$$ \Delta '\equiv -q ( q-q^{-1}) \mu \eta  q^N a_3 $$

Where $ N = N_1 + N_2 + N_3 $.
By using the expression ( 8 )  for $ \Delta $  and  $ \Delta' $  we find
the realizations of  $ a $  and $ k $ .

\begin{equation} k \equiv \mu^{-1} q^{-(N_1 + N_2 ) } ( a_3^{\dagger} +
q a_1^{\dagger}
a_2^{\dagger})\end{equation}
\begin{equation} a \equiv ( q-q^{-1} ) \lambda \mu \nu q^{N_3 } ( q
^{-2} a_3 + q^N ( q-q^{-1}) a_1
a_2 ) \end{equation}
By straightforward manipulations one can verify directly that the
elements $ a
, b , . . .  k  $ defined as above satisfy the commutation relations of the
quantum matrix algebra $ M_q (3) $.\\ \\
{\large \bf IV. Discussion}\\

Similar methods [10] have been used for studying the irreducible
* representations
of twisted $ SU(3) $ group [11]. However the  method of l
abeling the states and specially  the particular choice of raising and
lowering
operators that we have adopted ( see eq. (8-10) ) are completely different
from that of ref. [10]. With this choice the problem of classification of
all finite
dimensional irreducible representations of 3 by 3 quantum matrix groups
simplifies considerably.( see ref. [12] for the case of $ GL_q(3) $ ).
It may
be interesting to do the same thing for twisted $ SU(3) $.
\newpage

{\large \bf  References}
\begin{enumerate}
\item L. G. Biedenharn ; J. Phys. A. Math. Gen. 22 L873 (1989)
\item A. J. Macfarlane ; J. Phys. A. Math. Gen. 22 4551 (1989)
\item C. P. Sun and H. C. Fu ; J. Phys. A. Math. Gen. L983 (1989)
\item V. G. Drinfeld ; Proceeding of the ICM ( Berkeley, Berkeley , CA,
1986) p.798
\item M. Jimbo ; Lett. Math. Phsy. 10, 63 ( 1985) ; 11,247 (1986)
\item N. Yu. Reshetikhin , L. A. Takhtajan , and L. D.Faddeev ;
Alg. Anal. 1, 178 (
1989) in Russian
\item S. L. Woronowicz , Commun. Math. Phys. 111, 613 (1987); 130,387(1990)
\item Y. Manin CRM Preprint ( 1988)
\item M. L. Ge, X. F. Liu , and C. P. Sun : J. Math. Phys. 38 (7) 1992
\item K. Bragiel, Lett. Math. Phys. 17 (1989) 37-44
\item V. Karimipour ;Representations of the coordinate of $ GL_q(3) $ IPM
preprint 93-010 ,Tehran (1993) To Appear in Lett. Math. Phys. \end{enumerate}
\end{document}